\documentclass[epj]{svjour}
\usepackage{latexsym}
\usepackage{amsmath}
\usepackage{amssymb}
\usepackage{graphics}
\usepackage{graphicx}
\usepackage{tabularx}
\usepackage{geometry}
\usepackage{psfrag}
\usepackage{comment}
\usepackage{bm}
\usepackage{latexsym}
\usepackage{wasysym}
\usepackage{lineno}
\usepackage{dcolumn}
\usepackage{stackrel}

\makeatletter
\def\overlay#1#2{\mathpalette\@overlay{{#1}{#2}{\hfil}{\hfil}}}
\def\@overlay#1#2{\@@overlay#1#2}
\def\@@overlay#1#2#3#4#5{{%
   \def\overlaystyle{#1}%
   \setbox0=\hbox{\m@th$\overlaystyle#2$}%
   \setbox1=\hbox{\m@th$\overlaystyle#3$}%
   \ifdim \wd0<\wd1 \setbox2=\box1 \setbox1=\box0 \setbox0=\box2\fi
   \rlap{\hbox to\wd0{#4\box1\relax#5}}\box0%
}}
\makeatother

\newcommand{\SuperCirc}{\overlay{}{\circlearrowleft}}

\makeatletter
\newlength\Cover@height
\newlength\Cover@depth
\newlength\Cover@width
\newcommand{\Cover}[2][]{%
  \settoheight{\Cover@height}{$#2$}
  \settodepth{\Cover@depth}{$#2$}
  \settowidth{\Cover@width}{$#2$}
  \def\Cover@arg{#1}
  \raisebox{.5ex}{\raisebox{\Cover@height}{\rlap{%
  $\SuperCirc$}}}
  #2
}
\makeatother

\newcommand{\CirPol}{\Cover{\gamma}}

\begin{document}

\title{Polarized positron production at MeV electron accelerators}
%\subtitle{Do you have a subtitle?\\ If so, write it here}

\author{Eric Voutier\inst{1} 
{\thanks{\emph{Contact email:} voutier@ipno.in2p3.fr}} 
on behalf of the PEPPo Collaboration
}                     % Do not remove
\institute{Institut de Physique Nucl\'eaire, Universit\'es Paris-Sud \& Paris-Saclay, CNRS/IN2P3, 91406 Orsay, France}
\date{Received: date / Revised version: date}
\abstract{
The Polarized Electrons for Polarized Positrons (PEPPo) experiment has demonstrated the efficient transfer of polarization 
from electrons to positrons produced by the bremsstrahlung radiation of a polarized electron beam in a high-$Z$ target. 
Positron polarization up to 82\% has been measured for an initial electron beam momentum of 8.19~MeV/$c$, limited only 
by the electron beam polarization. Combined with the high intensity and high polarization performances of polarized 
electron sources, this technique extends efficient polarized positron capabilities from GeV to MeV electron accelerators. 
This presentation reviews the PEPPo proof-of-principle experiment and addresses the perspectives for future applications.
 % end of PACS codes
} %end of abstract

\authorrunning{E.~Voutier}
\titlerunning{Polarized positron production}

\maketitle

%
% --------------------------------------------------------------------------------------------------------------------------
%

\hyphenation{brems-strah-lung}
\hyphenation{po-si-tron}
\hyphenation{pro-ject}
\hyphenation{sour-ces}
\hyphenation{rea-ches}
\hyphenation{me-cha-nisms}
\hyphenation{lon-ger}
\hyphenation{de-mon-stra-ted}
\hyphenation{a-zi-mu-thal}
\hyphenation{so-le-noid}
\hyphenation{po-la-ri-me-ter}
\hyphenation{met-hod}
\hyphenation{de-du-ced}
\hyphenation{lon-gi-tu-di-nal}
\hyphenation{de-mon-stra-te}

\section{Introduction}
\label{intro}

The interest in both polarized and unpolarized positron beams for the experimental investigation of the physical world ranges from the macroscopic molecular scale accessible at eV energies down to the most elementary scale of fundamental symmetries probed with hundreds of GeV lepton beams. \newline
For energies up to a few hundred keV, they offer the unique opportunity to probe the surface magnetization of materials~\cite{Gid82} or to investigate their inner structural defects with unprecented resolution and accuracy~\cite{Kra99}. They are also of prime importance for the study of the Bose-Einstein condensate~\cite{Cas09}, and crucial for the development of anti-matter research~\cite{Cha17} and related new energy sources. \newline
In the GeV energy range where the electromagnetic interaction dominates lepton-hadron reactions, there is no stringent difference between the physics information obtained from the scattering of electrons or positrons off a nucleus target.  However, every time a reaction process is a conspiracy of more than one elementary mechanism, the comparison between electron and positron scatterings allows us to isolate the quantum interference between these mechanisms. This is of particular interest for studying limitations of the one-photon exchange Born approximation in elastic and inelastic scatterings, specifically effects of two-photon exchange mechanims~\cite{Gui03} which may reconciliate the differences between cross section and polarization measurements of 
the electric form factor of the proton~\cite{Pun15}. It is also essential for the experimental determination of the generalized parton distributions where the interference between the known Bethe-Heitler process and the unknown deeply virtual Compton  scattering requires polarized and unpolarized electron and positron beams for a model independent extraction~\cite{Vou14}. \newline
In the several tens to hundreds GeV energy range, the lepton-hadron interaction is no longer restricted to neutral 
current exchange and reaches the charged currents domain of the electroweak sector. Charged W$^{\pm}$ currents interact with electron and positron beams as essentially different experimental probes, able to uniquely isolate positively or negatively charged quarks. In the deep inelastic regime, charged current interactions access combinations of quark flavors different from those measured with purely electromagnetic neutral interaction, providing an alternative and novel source of information about parton distribution functions, particularly with respect to the strange and charm quarks. For instance, the availability of polarized electron and positron beams provide the necessary tools to measure the difference between the strange and anti-strange quark distributions free from any ambiguities related to the hadronization process~\cite{Asc13}. Such polarized lepton 
beams also provide the ability to test fundamental predictions of the Standard Model such as the absence of right-handed 
charged currents or some possible scenarios for the existence of a new physics beyond the frontiers of the Standard Model.

The production of high-quality polarized positron beams relevant to these many applications remains however a highly difficult task 
that, until recently, was feasible only at large scale accelerator facilities. Polarized positron beams based 
on radioactive sources are limited in polarization and intensity. The production of highly polarized and intense positron beams has been achieved at high energy storage facilities ($>10$~GeV) taking advantage of the self-polarizing Sokolov-Ternov effect~\cite{Sok64} for ultra-relativistic particles orbiting inside a magnetic field. Other scenarios~\cite{{Omo06},{Ale08}}  have been proposed within the context of the International Linear Collider project, but still involving high energy 
electron beams and challenging technologies that intrinsically limit their range of applications.

Relying on the most recent advances in high polarization and high intensity electron sources~\cite{Add10}, the PEPPo (Polarized Electrons for Polarized Positrons) technique~\cite{Gra11} provides a novel and potentially widely accessible approach based on the production, within a tungsten target, of polarized $e^+e^-$ pairs from the circularly polarized bremsstrahlung radiation of a low energy ($<$~10~MeV) highly polarized electron beam. This presentation reviews the principle of operation of this technique and its experimental demonstration performed at the injector of the Continuous Electron Beam Accelerator Facility (CEBAF) of the Thomas Jefferson National Accelerator Facility (JLab) in Newport News, VA. Further developments of the PEPPo method in the context of 
the JLab physics program and a possible low energy polarized positron beam facility are also sketched.
 
%
% --------------------------------------------------------------------------------------------------------------------------
%

\section{Polarized bremsstrahlung and pair creation}
\label{polprod}

Polarization phenomena in electromagnetic processes have been investigated since the early thirties~\cite{{Som31},{Wic51},{May51}}. The originally more complete calculations of the polarization of the bremsstrahlung radiation generated by an electron beam in the vicinity of a nuclear field~\cite{Ols59} drove the development of polarized photon beams: an upolarized electron beam is predicted to generate a linearly polarized photon beam, while a polarized electron beam would generate a circularly polarized photon beam with polarization directly proportionnal to the initial electron beam polarization. These features were used extensively at numerous accelerator facilities, and more recently in the experimental hall B of JLab~\cite{Mec03} to operate a high energy polarized photon beam. 

\begin{figure}[t]
\begin{center}
\resizebox{0.45\textwidth}{!}{\includegraphics{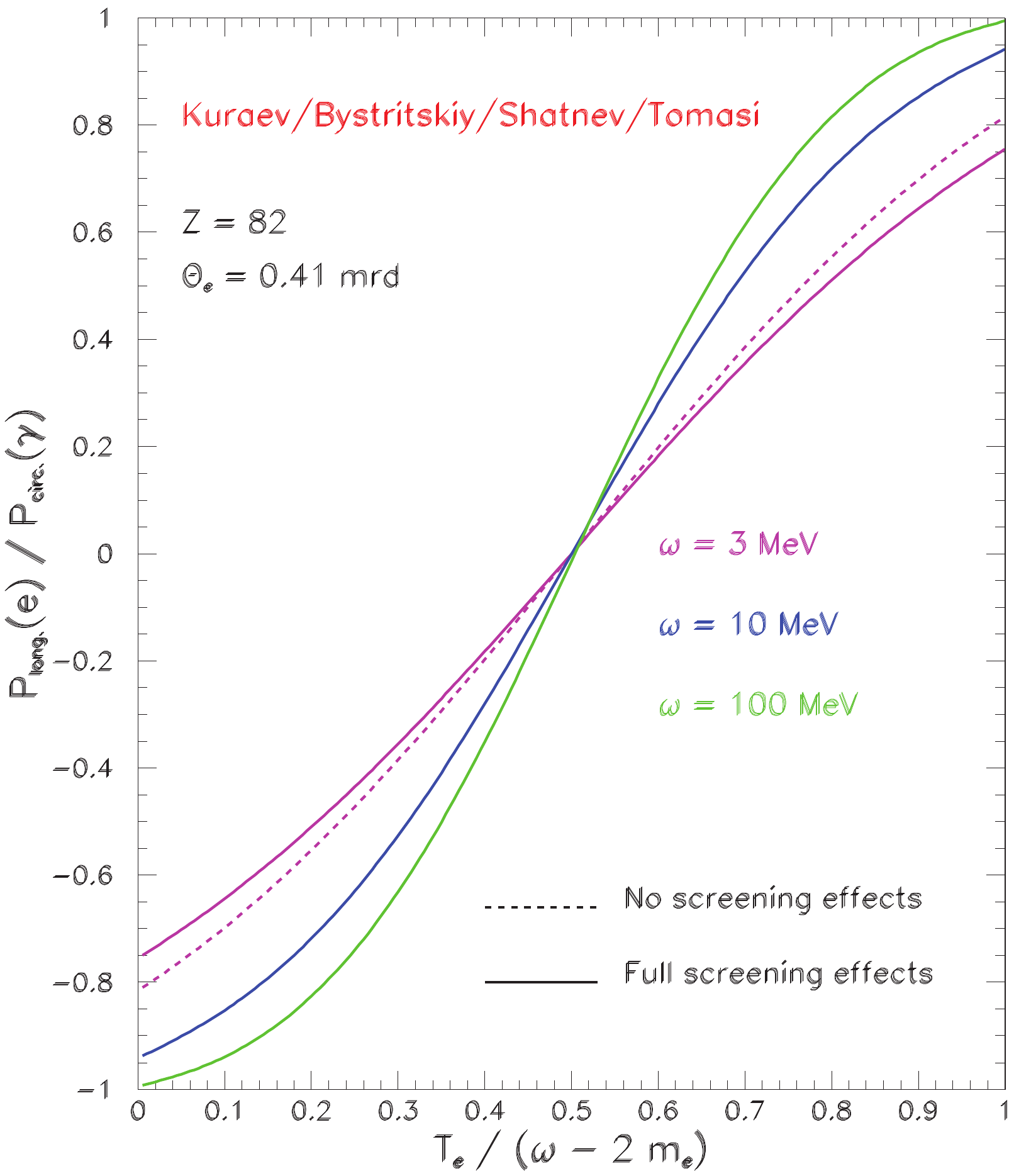}}
\caption{Circular-to-longitudinal polarization transfer from photons to positrons emitted at a small angle for different 
initial photon energies and within a high $Z$ material~\cite{Kur10}. Full lines indicate calculations within a full screening approach, and the dashed line corresponds to a calculation neglecting screening effects. Curves are represented as function of the full kinetic energy portion available to the positrons.}
\label{polpair}
\end{center}
\end{figure}

As a reciprocal process to bremsstrahlung, polarization observables of the pair production process can be obtained from the 
same expressions modulo some kinematical substitutions. Nevertheless, this recipe is of limited application since  ultra-relativistic approximations for the bremsstrahlung and pair production processes are somehow different in nature, 
especially close to the production threshold where lepton masses cannot be neglected~\cite{Dum09}. This was exactly 
demonstrated in a recent theoretical work where finite lepton mass effects were considered~\cite{Kur10}. Fig.~\ref{polpair} shows the circular-to-longitudinal polarization transfer calculated for a high $Z$ material and different initial beam conditions within the framework of Ref.~\cite{Kur10}. Depending on the kinetic energy of the created positrons, the polarization transfer  ranges from -1 to 1 with a remarkable kinematically symmetric behaviour. This quite natural feature for a final  system consisting of two particles identical in mass and spin, is absent from older calculations. As a consequence, the low energy half of the spectra  shows strong differences between ultra-relativistic and finite mass calculations, even when high initial photon energies are considered. These different approaches both predict that a circularly polarized photon beam creates a polarized $e^+e^-$ pair whose  longitudinal and transverse polarization components are both proportionnal to the initial photon beam polarization. Polarization transfer in the transverse plane is however much less efficient than in the longitudinal plane. The experimental demonstration of the circular-to-longitudinal polarization transfer is relatively recent and has been carried out at KEK~\cite{Omo06}, SLAC~\cite{Ale08} and JLab~\cite{Abb16} using completely different techniques to  produce the initial polarized photon beam.

%
% --------------------------------------------------------------------------------------------------------------------------
%

\section{The PEPPo experiment}
\label{peppo}

The technique used to produce circularly polarized photons is the essential difference between the new approaches for the production  of polarized positron beams at linear accelerator facilities. The concept demonstrated at KEK~\cite{Omo06} relies on the Compton 
back-scattering of polarized laser light off a $\sim$GeV electron beam to produce a roughly uniform photon spectra up to 
several tens of MeV. The undulator scheme demonstrated at SLAC~\cite{Ale08} involves the polarized synchroton radiation produced by a multi-GeV electron beam traveling within a helical undulator. Both these techniques are demanding in terms of the properties of the initial electron beam, consequently limiting their use to large scale facilities. On the contrary, the PEPPo  concept~\cite{{Bes96},{Pot97}} which relies on the bremsstrahlung radiation of a polarized electron beam can be used efficiently with a low energy ($\sim$5-100 MeV/$c$), high intensity ($\sim$mA), and high polarization ($>$~80\%) electron beam driver, providing  access to polarized positron beams to a much wider community because of the reduced cost of the driver accelerator. The PEPPo  experiment~\cite{Gra11} was designed to evaluate the performance of the PEPPo concept by measuring the polarization transfer from a primary electron beam to the produced positrons. 

\subsection{Principle of operation}
\label{peppo-1}

\begin{figure}[t]
\begin{center}
\includegraphics[width=0.95\linewidth,angle=0]{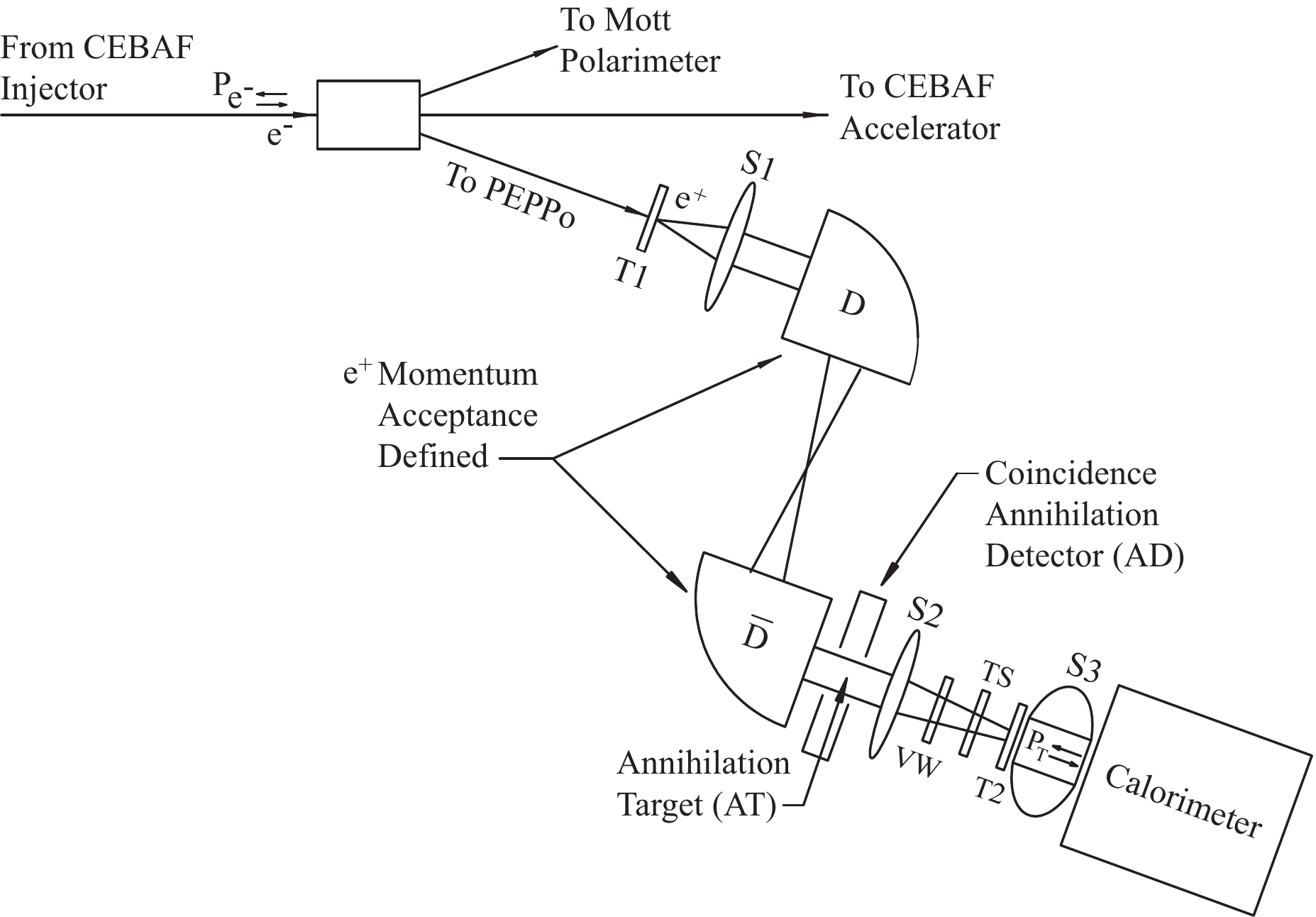}
\caption{\label{PEPPo-line} Schematic of the PEPPo line and apparatus illustrating the principle of operation of the experiment 
based on the processes sequence $\protect\overrightarrow{e^-} \stackrel{\mbox{\tiny{T1}}}{\rightarrow} \CirPol 
\stackrel{\mbox{\tiny{T1}}}{\rightarrow} \protect\overrightarrow{e^+} \stackrel{\mbox{\tiny{T2}}}{\rightarrow} \CirPol 
\stackrel{\mbox{\tiny{S3}}}{\rightarrow} \gamma$ described in the text. The setup footprint is about 3$\times$1.5~m$^2$.}
\end{center}
\end{figure}

The PEPPo experiment ran at the CEBAF injector~\cite{Kaz04} of JLab in june 2012 using a polarized electron beam up to  $p_e$=8.19$\pm$0.04~MeV/$c$ to measure the momentum dependence of the polarization of the produced positrons in the momentum 
range 3.07-6.25~MeV/$c$. \newline A new beam line (Fig.~\ref{PEPPo-line})~\cite{Vou13} was constructed where polarized electrons were transported to a 1~mm thick tungsten positron production target (T1) followed by a positron collection, selection, and characterization system~\cite{Ale09}. Longitudinally polarized electrons interacting in T1 radiate elliptically polarized photons whose circular component ($P_{\gamma}$) is proportional to the electron beam polarization ($P_e$). Within the same target, the polarized photons produce polarized $e^+e^-$-pairs with perpendicular ($P_{\perp}$) and longitudinal ($P_{\parallel}$) polarization components both proportional to $P_{\gamma}$ and therefore $P_e$. The azimuthal symmetry causes $P_{\perp}$ to vanish resulting in longitudinally polarized secondary positrons. Immediately after T1, a short focal length solenoid (S1) collects the positrons into a combined function spectrometer ($\mathrm{D}\overline{\mathrm{D}}$) that uses two 90$^{\circ}$ dipoles and a slit to select positron momentum. The exiting positrons can either be detected at a positron diagnostic (AT+AD) or refocused by a second solenoid (S2) through a vacuum window (VW) to a Compton transmission polarimeter. Retracting T1, the known electron beam could also be transported to T2 to calibrate the polarimeter analyzing power. 

\noindent
The polarimeter~\cite{Ale09} begins with a 2~mm densimet (90.5\%W/7\%Ni/2.5\%Cu) conversion target (T2) followed by a 7.5~cm 
long, 5~cm diameter iron cylinder centered in a solenoid (S3) that saturates and polarizes it (creating a known polarization,  $\overline{P_T}$). An electromagnetic calorimeter with 9 CsI crystals arranged in a 3$\times$3-array is placed at the exit of 
the polarimeter solenoid. Polarized positrons convert at T2 via bremsstrahlung and annihilation processes into polarized photons which  polarization orientation and magnitude depend on the positron polarization. As a consequence of the polarization dependence 
of the Compton process, the number of photons passing through the iron core and subsequently detected by the CsI-array depends on the relative orientation of the photon and iron core polarizations. By reversing the sign of the positron polarization (that is reversing the electron beam helicity) or the target polarization (via S3 polarity), one measures the experimental Compton asymmetry 
\begin{equation}
A_C^p = P_{\parallel} \, \overline{P_T} \, A_p = \epsilon_P \, P_e \, \overline{P_T} \, A_p \label{eq:ComAs}
\end{equation}
where $A_p$ is the positron analyzing power of the polarimeter and $\epsilon_P$ is the electron-to-positron polarization transfer efficiency. Knowing $P_T$, $P_e$, $A_p$ and measuring $A_C^p$ provide a measurement of $P_{\parallel}$ and consequently $\epsilon_P$.

\subsection{Calibration}
\label{peppo-2}

The complete experiment was first calibrated with the known CEBAF electron beam at each selected positron momentum. \newline 
The accelerating gradient in the cavities of the quarter-cryomodule of the CEBAF injector was adjusted by determining the magnitude of the magnetic field required to deflect the beam towards a specific location~\cite{Kaz04}, providing a 0.7\% measurement of the beam momentum. The same field with opposite polarity deflects the beam into the PEPPo line towards the production target. \newline
In absence of T1, transporting the beam to a Faraday cup (FC) at the end of the PEPPo line allows us to set the  $\mathrm{D}\overline{\mathrm{D}}$ spectrometer current. This procedure was repeated for each of the positron momenta studied. 
Using the nominal 8.19~MeV/$c$ momentum of the PEPPo experiment and inserting T1 in the beam path allowed us to establish the S1 current by optimizing the transported flux at FC for each momentum. Further optimization of that same flux while powering S2 set the current of that last magnet of the PEPPo beam transport line. Reversing the $\mathrm{D}\overline{\mathrm{D}}$ polarity is selecting and transporting to T2 positrons produced at T1. T2 is mounted on an electrically isolated aluminum target holder (TH) connected to a picoammeter and serving as a relative positron flux monitor. The S1 and S2 current optimization procedure was repeated with respect to the TH beam diagnostic for positive $\mathrm{D}\overline{\mathrm{D}}$ polarity, and found to be in good agreement with previously optimized current values, supporting a negligible effect of the polarimeter magnet on incoming particle trajectories. The experimentally determined S1, $\mathrm{D}\overline{\mathrm{D}}$, and S2 currents agree well with those determined by a GEANT4~\cite{Ago03} model of the experiment using magnetic fields modeled with OPERA-3D~\cite{Ben15}. 

Following Eq.~\ref{eq:ComAs}, the measurement of the Compton asymmetry for a known beam permits the calibration of the analyzing power 
of the polarimeter~\cite{Ade13} assuming knowledge of the beam and target polarizations. \newline  
The polarization of the electron beam, $P_e$, was measured to be $85.2\pm0.6\pm0.7$\% with a Mott polarimeter~\cite{Gra13}. The 
first uncertainty is statistical and the second is the total systematic uncertainty associated with the theoretical and experimental 
determination of the Mott analyzing power. \newline
The target polarization was experimentally deduced from the measurement of the magnetic flux induced in a set of pick-up coils surrounding the iron core~\cite{Ale09}, that was generated when powering the S3 magnet to the nominal 60~A operational current~\cite{Dad13}.  The comparison of the experimental signal with the simulated signal modeled with OPERA-3D provides the average longitudinal magnetic field polarizing the iron core $\overline{B_z} = 2.935 \pm 0.036$~T~\cite{Fro14}. The corresponding average longitudinal polarization was determined to be $\overline{P_T} = 7.06\pm0.09$\%~\cite{Fro14}, in very good agreement with the previously reported value~\cite{Ale09}.

\subsection{PEPPo measurements}
\label{peppo-3}

\subsubsection{Data recording}

Electrons or positrons arriving at T2 convert into photons that eventually fire the crystal array. The signal delivered by each crystal is read with a R6236-100 Hamamatsu photomultiplier (PMT). The effective gain of the full electronics chain of each crystal was  calibrated prior beam exposure with $^{137}$Cs and $^{22}$Na radioactive sources, and monitored throughout data taking by controlling the position of the 511~keV peak produced by the annihilation of positrons generated inside the core. This method insures a robust and stable energy measurement, intrinsically corrected for possible radiation damages or PMT-aging effects. \newline
The PMT signals are fed into an analog-to-digital converter module that samples the signal into 500 successive time bins of 4~ns duration and enables three data taking modes. The sample mode, used for detector monitoring, allows registration of the full sample set whenever an external trigger is  provided. The semi-integrated mode, used for positron data taking, collapses the 500 samples into 1 integrated value registered for each positron trigger. The latter is built from the coincidence between the central crystal and a 1~mm thick scintillator (TS on Fig.~\ref{PEPPo-line}) placed between the PEPPo line vacuum exit window and T2; it provides an  effective charged particle trigger that considerably reduces the neutral backgrounds into the crystal array. The last or integrated data taking mode integrates the signal delivered by the crystal over a duration corresponding to a fixed beam polarization orientation (helicity gate). It is specific to high rate, background-free environments, and was used for the electron calibration measurements.

\subsubsection{Experimental asymmetries}

The comparison of the energy deposit ($E^{\pm}_{\lambda h}$) during the helicity gate for each beam polarization orientation and fixed analyzing magnet polarity ($\lambda$=$\pm 1$) gives the experimental asymmetry
\begin{equation}
A_C^e (\lambda,h) = \frac{E^{+}_{\lambda h}-E^{-}_{\lambda h}}{E^{+}_{\lambda h}+E^{-}_{\lambda h}}  
\end{equation}
where $h$=$\pm 1$ indicates the beam helicity status at the electron source, that can be reversed inserting a half-wave plate in the 
excitation laser light pathway~\cite{Kaz04}. Data taking was repeated for each magnet polarity and beam helicity, and were 
statistically combined 
\begin{eqnarray}
A_C^e & = & \sum_{\lambda h}  \frac{A_C^e (\lambda,h)}{\left( \delta A_C^e (\lambda,h) \right)^2} \bigg{/} \sum_{\lambda h}  
\frac{1}{\left( \delta A_C^e (\lambda,h) \right)^2} \nonumber \\
& = & P_e \, \overline{P_T} \, A_e \label{eq:StaC} 
\end{eqnarray}
to provide the actual Compton asymmetries $A_C^e$ for electrons, free from false asymmetries related to the beam or the analyzing magnet. $A_e$ represents here the electron analyzing power of the polarimeter. Electron experimental data feature high 
statistical accuracy ($<1$\%) and similar systematic errors originating from the determination of the pedestal signal.

Positron data are recorded on an event-by-event basis and, because of the trigger configuration, involve only the central crystal. The experimental information consists of the energy deposit in the crystal and the coincidence time ($t_c$) with TS. For each helicity gate ($i$) and $\lambda h$ configuration, the energy yield $Y^{\pm}_{i, \lambda h}$ is constructed 
\begin{equation}
Y^{\pm}_{i, \lambda h} = \sum_j \frac{ N^{\pm}_{t,ij} \, \mathcal{E}_{ij} }{Q^{\pm}_{i} \, dt^{\pm}_{i}}
\end{equation}
where the sum runs over the events occuring during the helicity gate; $\mathcal{E}_{ij}$ is the energy deposit in the crsytal, $Q_i^{\pm}$ is the helicity gate beam charge determined from a beam current monitor on the main accelerator line, and $dt^{\pm}_{i}$ represents the electronics and data acquisition dead-time correction measured with specific helicity gated scalers. $N^{\pm}_{t,ij}$ is the true number of coincidence events within a selected time window around the $t_c$ peak, determined from 
\begin{equation}
N^{\pm}_{t,ij} (\lambda,h) = N^{\pm}_{m,ij} (\lambda,h) - N^{\pm}_{r,ij}
\end{equation}
where $N^{\pm}_{m,ij}$ is the measured number of events within the selected time window and $N^{\pm}_{r,ij}$ is the random coincidences  contamination deduced from the fit of the global $t_c$ spectra. Positron experimental asymmetries are obtained analogously to electron asymmetries by combining each $\lambda h$ configurations   
\begin{equation}
A_C^p (\lambda,h) = \frac{\sum_i Y^{+}_{i,\lambda h} - \sum_i Y^{-}_{i,\lambda h}}{\sum_i Y^{+}_{i,\lambda h} + \sum_i Y^{-}_{i,\lambda h}} 
\end{equation}
according to 
\begin{eqnarray}
A_C^p & = & \sum_{\lambda h}  \frac{A_C^p (\lambda,h)}{\left( \delta A_C^p (\lambda,h) \right)^2} \bigg{/} \sum_{\lambda h}  
\frac{1}{\left( \delta A_C^p (\lambda,h) \right)^2} \nonumber \\
& = & P_{\parallel} \, \overline{P_T} \, A_p \, .
\end{eqnarray}
Experimental asymmetries and uncertainties for each positron momentum are obtained by insuring a mimimum energy deposit $\mathcal{E}_{ij} > 511$~keV. Main sources of systematics originate from the energy calibration procedure, the random subtraction  method, and the selection of coincidence events. The relative contribution to asymmetry uncertainties does not exceed 2\%.   

\subsubsection{Polarimeter analyzing power}

\begin{figure}[t!]
\includegraphics[width=0.985\linewidth]{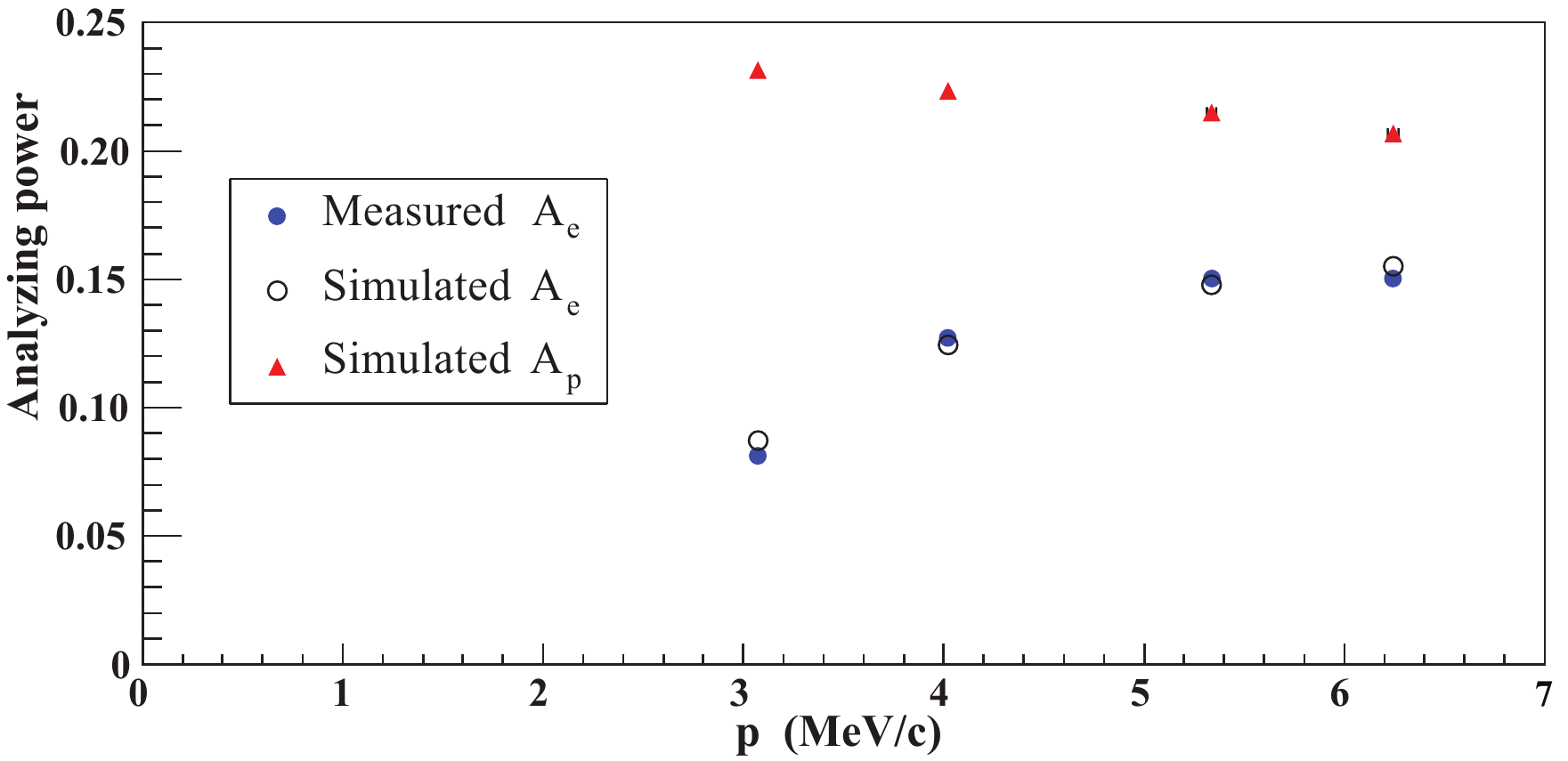}
\caption{\label{AnaPow} Electron and positron analyzing powers of the central crystal of the polarimeter. 
Statistical uncertainties were combined quadratically with systematic uncertainties taken from $P_e$, $\overline{P_T}$, and 
$A_C^e$ to determine acutal error bars~\cite{Abb16}.}
\end{figure}

The PEPPo beam line, magnet fields, and detection system was modeled using GEANT4, taking advantage of a previous implementation of polarized electromagnetic processes~\cite{Dol06}. The determination of the analyzing power of the polarimeter for positrons relies on 
the comparison between experimental and simulated electron asymmetries. This benchmarks the GEANT4 physics packages and resolves related  systematic uncertainties within the limits of the measurement accuracy. The measured electron analyzing power is shown in Fig.~\ref{AnaPow} and compared with the predictions of the GEANT4 model of the experiment. The reamarkable agreement between data and simulations indicates an accurate understanding of the full experimental setup. The analyzing power of the polarimeter for 
positrons (Fig.~\ref{AnaPow}) is then directly simulated. The combination of the supplementary $e^+$-to-$\gamma$ annihilation conversion process together with the minimum energy deposit requirement leads to larger analyzing power for positrons than electrons. The energy 
cut effect is strong at low $e^+$ momenta where it removes a significant part of the energy spectra acting as a dilution of the polarization sensitivity. 

\subsubsection{Positron polarization}

\begin{figure}[t!]
\begin{center}
\includegraphics[width=0.975\linewidth]{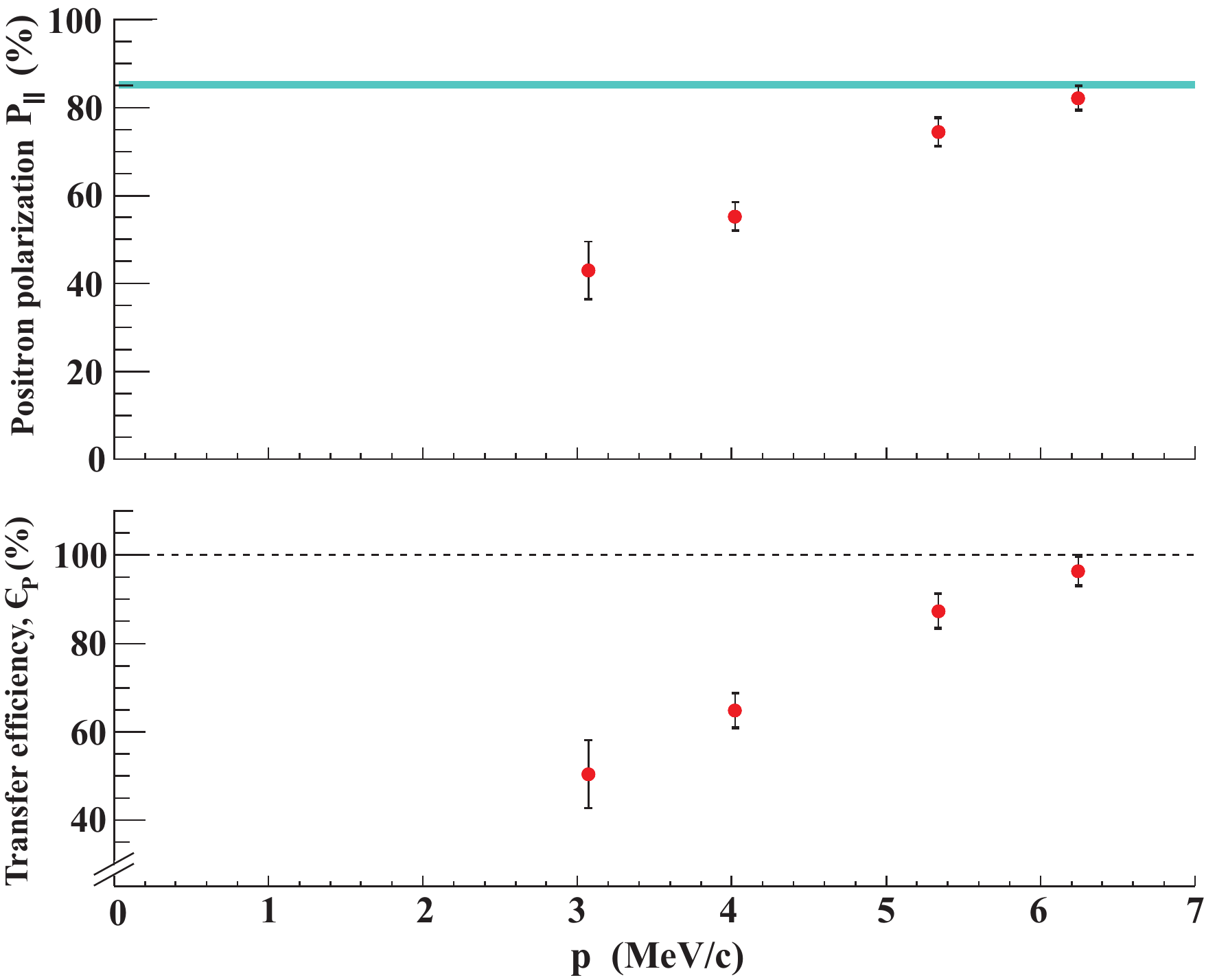}
\caption{\label{PolPos} PEPPo measurements of the positron polarization (top panel) and polarization transfer 
efficiency (bottom panel); statistics and systematics are reported for each point, and the shaded area indicates the electron 
beam polarization~\cite{Abb16}.}
\end{center}
\end{figure}

The positron longitudinal polarization $P_{\parallel}$ and the polarization transfer efficiency $\epsilon_P$ are obtained from 
Eq.~\ref{eq:ComAs} knowing the electron beam polarization, the target polarization, and the positron analyzing power. The PEPPo 
data demonstrate a remarkably highly efficient polarization transfer from photons to positrons over 
a large positron momentum range (Fig.~\ref{PolPos}). The high polarization observed at energies significantly smaller than the kinematically achievable maximum is most likely a consequence of multiple scattering effects: higher energy particles with higher polarization 
suffer loss of energy when interacting with atomic electrons but their polarization remains globally unchanged. The 
bremsstrahlung of longitudinally polarized electrons is therefore demonstrated to be an efficient process for generating longitudinally 
polarized positrons. 

%
% --------------------------------------------------------------------------------------------------------------------------
%

\section{Polarized positron source perspectives}
\label{persp}

Beyond \, the \, polarization \, performance \, of the PEPPo technique, one very appealing feature is the low energy of the electron 
beam driver. This does not percludes application of this method at high energies but signficantly reduces the cost and 
technological requirements for developing a polarized positron beams. In addition, the radiation environment of low energy 
sources is easier to handle, making the PEPPo concept easier to realize. It is therefore more realistic to 
conceive of application of this technique in many different research fields.

\begin{figure}[th]
\begin{center}
\includegraphics[width=0.985\linewidth]{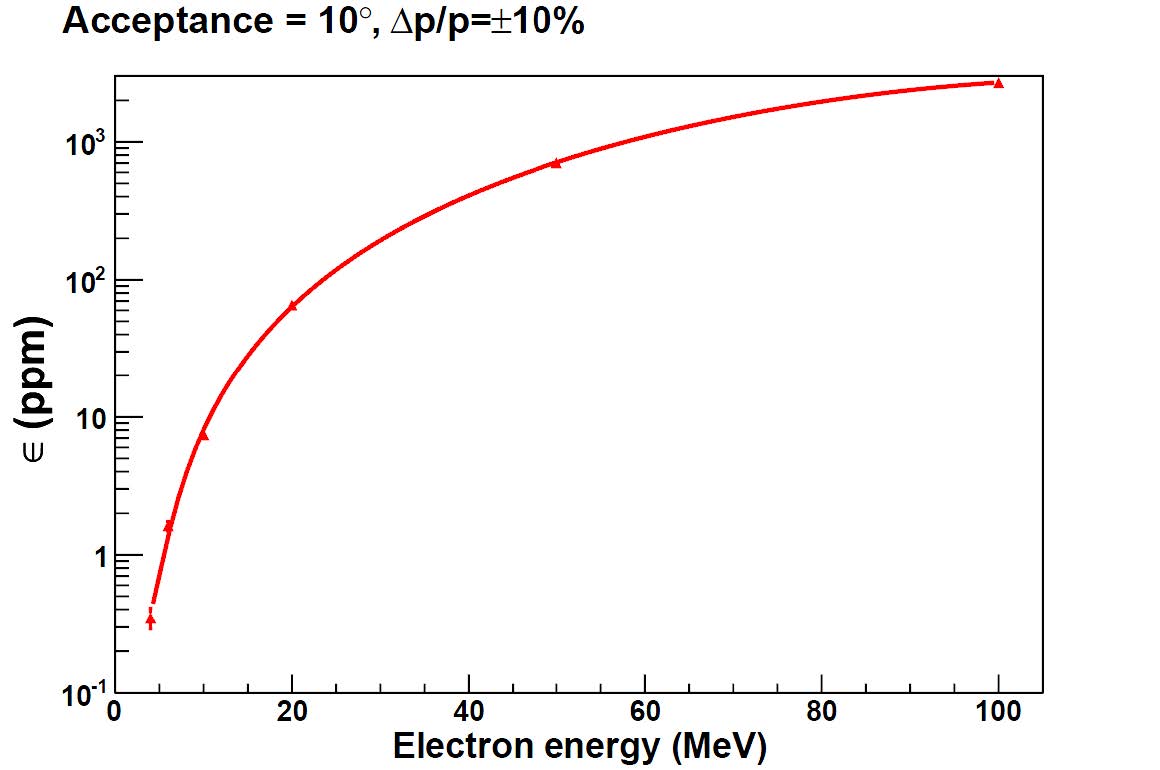}
\includegraphics[width=0.985\linewidth]{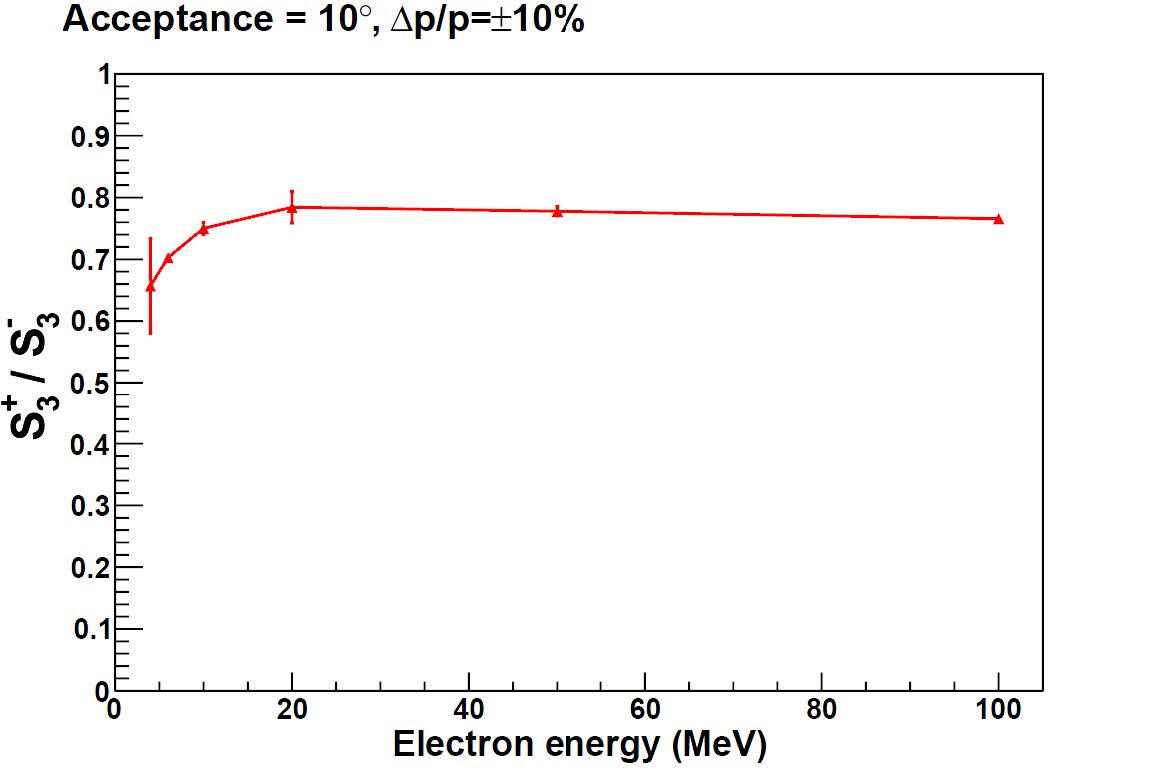}
\caption{\label{PoSour} GEANT4 simulations of the beam energy dependence of the positron production efficiency (top panel) and the circular-to-longitudinal polarization transfer (bottom panel) at the positron energy corresponding to the optimum FoM~\cite{Dum11}.}
\end{center}
\end{figure}

As a general benefit for any final use, the present technological research is directed towards maximizing the positron flux 
both by increasing the initial electron flux and by enhancing the positron collection system. It is worth noting that all three of the 
demonstrated techniques based on circularly polarized photons operate within a similar positron energy range. \newline
Recently, polarized electron sources have demonstrated sustainable operation at 1~mA and various scenarios 
can be developed to accumulate 10~MeV electrons and increase further the initial electron flux before the production target. The power constraint on this target is very strong, requiring operation at $\sim$10~kW in the lowest intensity case. Potential targets have already been proposed; in particular a cooled liquid metal target developed at Niowave Inc.~\cite{NioIn} may solve this problem. \newline
The optimization of the collection system is a sophisticated problem requiring many variables to be specified for an optimal solution, particularly the positron energy range and the type of applications. The driving parameter for a polarized source is 
the Figure-of-Merit (FoM) which is the product of the positron polarization squared and beam intensity. As a rule of thumb, the optimum FoM of the PEPPo technique is obtained at half of the energy of the incoming electron beam, defining the optimum positron energy to be collected. This is a consequence of the fact that in bremsstrahlung based generation of positrons, smaller positron energies correspond to larger flux but lower polarization while higher positron energies lead to lower flux but higher polarization. \newline
The magnitude of the positron flux or efficiency of the production scheme ($\epsilon$ on Fig.~\ref{PoSour}) is essentially controlled by the accepted phase space for the positrons. Fig.~\ref{PoSour} shows the beam energy evolution of the optimized efficiency (top panel) and of the electron-to-positron polarization transfer (bottom panel) for a fixed positron acceptance as obtained from GEANT4 simulations~\cite{Dum11}. The saturation behaviour of $\epsilon$ should be noted. It results from the increase of both the pair creation cross section and the positron optimum energy as function of the electron beam energy. It clearly indicates that the gain 
in the positron production rate is limited beyond $\sim$100~MeV. At the positron energy corresponding to the optimized FoM, the polarization transfer is less sensitive to the initial beam energy. Over the 10-100~MeV electron energy range, one may expect to obtain an optimized polarized positron production with a flux about 10$^{-5}$-10$^{-3}$~e$^+$/e$^-$ and 75\% of the initial electron beam polarization. The exact final numbers depends on the constraints of the designated applications, mainly the positron momentum acceptance for accelerator applications and the energy moderation process for low-energy positron beams. In the latter case, it will be more efficient to limit the electron energy range to about 10~MeV and develop deceleration methods~\cite{Che09} to reach desired low-energies for the positrons. However, the implementation of these techniques remains a delicate task because of the intricate relationship between the geometry of the full system, the energy distribution of the particles to be decelerated and the phase of 
the electric field~\cite{Ang17}. The limitation of the transverse size of the generated positron beam without significant flux loss is an additional issue to resolve. 
   
%
% ------------------------------------------------------------------------------------------------------------------------
%
\section{Summary}
\label{summ}

The PEPPo experiment demonstrated a new concept for creating polarized positron beams by using low energy ($\sim$10~MeV) highly 
polarized electron beams. The measured 82~\% high positron polarization is limited only by the initial electron polarization. 
Considering the low energy range of the electron beam driver and the modern capabilities of polarized electron sources, we 
believe that this new concept is of interest for a wide community ranging from atomic and condensed matter physics up to nuclear 
and particle physics. 

%
% ------------------------------------------------------------------------------------------------------------------------
%

\begin{acknowledgement}

We are deeply grateful to the SLAC E-166 Collaboration, particularly K.~Laihem, K.~McDonald, S.~Riemann, A.~Sch\"alicke, P.~Sch\"uler, J.~Sheppard and A.~Stahl for loan of fundamental equipment parts and support in GEANT4 modeling. We also thank N.~Smirnov for delivery of critical hardware. This work was supported in part by the U.S. Department of Energy, the French  Centre National de la Recherche Scientifique, and the International Linear Collider project. Jefferson Science Associa-tes operates the Thomas Jefferson National Accelerator Facility under DOE contract DE-AC05-06OR23177.

\end{acknowledgement}

%
% --------------------------------------------------------------------------------------------------------------------------
%

%
% --------------------------------------------------------------------------------------------------------------------------
%

\end{document}